# Running CMS software on GRID Testbeds

D. Bonacorsi, P. Capiluppi, A. Fanfani, C. Grandi
*Universita' di Bologna and INFN Sezione di Bologna, Bologna, Italy*

M. Corvo, F. Fanzago, M. Sgaravatto, M. Verlato
*INFN Sezione di Padova, Padova, Italy*

C. Charlot, I. Semeniuok
*Laboratoire Leprince-Ringuet, Ecole Polytechnique, IN2P3, Palaiseau, France*

D. Colling, B. MacEvoy, H. Tallini
*Imperial College London, London, United Kingdom*

M. Biasotto, S. Fantinel
*Laboratori Nazionali di Legnaro dell'INFN, Legnaro, Italy*

E. Leonardi, A. Sciaba'
*CERN and INFN, Geneva, Switzerland*

O. Maroney
*University of Bristol, Bristol, United Kingdom*

I. Augustin, E. Laure, M. Schulz, H. Stockinger, V. Lefebure
*CERN, Geneva, Switzerland*

S. Burke
*Rutherford Appleton Laboratory (RAL), Didcot, United Kingdom*

J-J. Blaising
*Laboratoire d''Annecy-le-Vieux de Physique des Particules (LAPP), IN2P3, Annecy-le-Vieux, France*

J. Templon
*NIKHEF, Amsterdam, Holland*

M. Reale
*INFN CNAF, Bologna, Italy*

G. Tortone
*INFN Sezione di Napoli, Napoli, Italy*

F. Prelz
*INFN Sezione di Milano, Milano, Italy*

C. Loomis
*Laboratoire de l''Accelerateur Lineaire (LAL), IN2P3, Universite de Paris-Sud (ParisXI), France*

Starting in the middle of November 2002, the CMS experiment undertook an evaluation of the European DataGrid Project (EDG) middleware using its event simulation programs. A joint CMS-EDG task force performed a "stress test" by submitting a large number of jobs to many distributed sites. The EDG testbed was complemented with additional CMS-dedicated resources. A total of ~ 10000 jobs consisting of two different computational types were submitted from four different locations in Europe over a period of about one month. Nine sites were active, providing integrated resources of more than 500 CPUs and about 5 TB of disk space (with the additional use of two Mass Storage Systems). Descriptions of the adopted procedures, the problems encountered and the corresponding solutions are reported. Results and evaluations of the test, both from the CMS and the EDG perspectives, are described.

## 1. INTRODUCTION

The Compact Muon Solenoid experiment (CMS) [1] is one of the four particle physics experiments that will collect data at the Large Hadron Collider (LHC) [2] being built at CERN (Geneva, Switzerland) [3].

While the CMS detector will not begin taking data until 2007, hundreds of physicist around the world, members of the CMS collaboration, are currently taking part in compute-intensive Monte Carlo simulation studies of the detector and its potential for uncovering new physics.

The challenge for the CMS computing infrastructure is therefore to cope with the very large computational and data access requirements. The size of the resources required, the complexity of the software and the physical distribution of the CMS collaboration naturally imply a distributed computing and data access solution.

The Grid paradigm is one of the most promising solutions to be investigated, and CMS is collaborating with many Grid projects around the world in order to explore the maturity and availability of middleware implementations and architectures.

CMS decided to actively participate in the Grid projects since their outsets, with the aim of understanding how the Grid can be useful for CMS and how CMS software needs to be adapted in order to maximize the benefit of using Grid functionality and tools.

The current CMS Monte Carlo based programs (known as "CMS Production and Analysis") [4] were run on Grid testbed implementations, to provide "real-life" evaluation of the readiness and usability of currently delivered Grid middleware.

The European DataGrid project (EDG) [5] is a three-year EU funded program under the V framework.

The test described here (referred to as a "Stress Test" because of its high demand on the availability and responsiveness of software and hardware resources) evaluated the EDG middleware deployed on the EDG testbed in its second year of implementation of Grid functionality. The main goals were defined as:
- Verify the robustness of the Grid middleware in a production environment and provide feedback for CMS software design;
- Manage effectively the dynamic addition and removal of heterogeneous institutional resources in the "CMS production" environment;
- Produce data for physics studies of CMS, possibly at the level of million simulated events.

Some description of the testbed and the results obtained are discussed in this document. More information can be found in [6].

## 2. CMS SOFTWARE FOR MONTE CARLO PRODUCTION

CMS Monte Carlo production consists pipelining several stages together where the output of one stage serves as the input to the next. The longest stages are typically CPU-bound, but some are I/O-bound, and some vary depending on the data to be processed. Eventually some stages can be performed in a single step, thus avoiding the partial recording of intermediate results; this "step grouping" process is only possible for particular studies, when the intermediate steps do not require other external input.

The typical CMS Monte Carlo production jobs were:
- CMKIN, generation of the physical process to be simulated (job input is a simple set of generator parameters, output is a file named "ntuples");
- CMSIM, simulation of the CMS detector and particle behaviors (input is the file(s) generated by CMKIN with the addition of some control parameter, output is a file named "fz file");
- ORCA, reconstruction of CMS detector response and physics object creation (input is the pre-processed CMSIM output by a object digitization program interfaced with Objectivity/DB [7], output is a collection on an Objectivity/DB);
- "Ntuple only", the nickname to identify all the previous steps done in a single pass with the addition of a final process producing data directly usable by the analysis (input is all the required parameters and output is the final "ntuple file" used for analysis).

Table 1 gives some of the computational characteristics of the quoted CMS production stages.

Since Objectivity/DB was not deployed on the EDG testbed, in the Stress Test only the first two steps of the chain were tested. Each job used for the tests had to process 125 events.

Table 1: Size of data samples and CPU time per event simulation of the different CMS production stages for a typical physics channel production.

| MC production stage | Size/event | Time/event (PIII 1 GHz CPU) |
|---|---|---|
| **CMKIN** | ~ 0.05 MB (Ntuple) | ~ 0.4-0.5 sec |
| **CMSIM** | ~ 1.8 MB (Fz file) | ~ 6 min |
| **ORCA** | ~ 1.5 MB (Objy DB) | ~ 18 sec |
| **Ntuple "only"** | ~ 0.001 MB (Ntuple) | ~ 380 sec |

The main distinguishing feature of the CMS Monte Carlo production environment is that it involves production processing on a large scale while at the same time minimizing the amount of direct human intervention. It has automated subsystems dealing with the following areas: input parameter management, robust and distributed request and production accounting, preparation of executables, management of production resources, local access to mass storage, and distributed file storage and replica management.

All Monte Carlo production requests are stored in a reference database (*RefDB* [8]) at CERN. Each request contains all the input parameters needed to create the data. The request is dispatched to Regional Centers by e-mail. A set of scripts (IMPALA [4]) has been developed to automate production job creation and submission for the different steps of the production chain in a Regional Center. BOSS [9] is a system that is able to perform bookkeeping of the relevant information produced by the different types of jobs synchronously with job execution. The summary of job tracking performed by IMPALA using BOSS is sent back to the *RefDB*.

More recently, a Python based package, MCRunjob [10] was developed, providing a metadata based approach for specifying more complex workflow patterns, translating them into a set of submittable jobs in a variety of environments (including the legacy IMPALA one). MCRunjob is also able to "chain" the production steps into a single job.

More details of the architecture of the production machinery can be found in the extensive Spring 2002 DAQ TDR Production note [4].

## 3. CMS GRID APPROACH

The strategy to approach and integrate the Grid paradigm in the CMS software (production and analysis) is also described in [11].

Tests presented in this paper are the first early large-scale CMS trials of a Grid-aware environment for real-life applications, aiming for results indicating the usability of the middleware implementations. The same tests were planned to give a possible feedback to modify CMS software for adaptation to the Grid tools.

The CMS "Stress Test" activity on the EDG Testbed had in particular three main goals:
- Verification of the portability of the CMS production environment into a grid environment;
- Verification of the robustness of the European DataGrid middleware in a production environment;
- Production of data for the Physics studies of CMS, with an ambitious goal of ~ 1 million simulated events in a 5 weeks time.

Detailed measurements of performances and identification of possible bottlenecks were therefore planned also before the actual start of the test.

The test used as much as possible of the EDG middleware provided functionalities, which are positioned at high level in the Grid "layered model".

Tested functionalities included the Workload Management System (Resource Broker in particular), the Data Management System (Replica Manager and Replica Catalog in particular), the Globus [12] Information System (MDS) implemented by EDG and the Virtual Organization Management System (the way to manage user authorization and authentication within a community of persons with similar scientific interests). Other accessory functionalities were also planned and tested, as e.g. the logging and monitoring systems.

This approach can be considered as a "top-down" approach, as it tests the layered functionalities of the Grid middleware from "above". To perform these tests it was necessary to adapt or modify the CMS production tools in order to develop custom access APIs (or interfaces) to the "high" middleware provided components.

## 4. EUROPEAN DATAGRID (EDG) TESTBED

The testbeds are key elements of the EDG Project program as they provide the verification of usability of the Grid middleware by many users.

Many evolving and dynamically configured testbeds were foreseen and deployed. The "Application Testbed" is the one dedicated to prove the readiness for use by the current Applications' software that participates to the EDG Project. This testbed was extensively tested ("stress test") to measure the performances and the eventual feedback to Grid developers and CMS software designers.

Each main partner of the EDG EU project was committed to deploy a main site for tests. Sites were requested to deploy and dynamically maintain all the necessary grid services.

Main EDG testbed sites were (all EU Tier1 classified sites): CERN/Geneva (CH), CNAF/Bologna (IT), CC-IN2P3/Lyon (FR), NIKHEF/Amsterdam (NL), and RAL/Oxford (UK).

Additional CMS sites and resources (mostly EU Tier2 sites) were added dynamically to the testbed, in order to increase the total available resources and also to test the easiness of site participations.

Table 2 summarizes the utilized resources for the EDG CMS Stress Test.

Table 2: Sites and resources of EDG Testbed. The CMS added sites are marked with a "*". MSS indicates the presence and use of a Mass Storage system (Tape Robot).

| Site | Number of CPUs | Disk Space GB | Availability of MSS |
|---|---|---|---|
| CERN (CH) | 122 | 1000* (+100) | yes |
| CNAF (IT) | 40* | 1000* | |
| RAL (UK) | 16 | 360 | |
| Lyon (FR) | 120 (400) | 200 | yes |
| NIKHEF (NL) | 22 | 35 | |
| Legnaro (IT)* | 50 | 1000* | |
| Ecole Polytechnique (FR)* | 4 | 220 | |
| Imperial College (UK)* | 16 | 450 | |
| Padova (IT)* | 12 | 680 | |
| Totals | 402 (400) | 3000* +(2245) | |

The test run from November 30th 2002 to Xmas 2002, lasting therefore about three weeks time.

EDG organization included as a key part of the Project the participation of "Applications" representatives to drive the middleware requirements and the testing of the implemented solutions. This approach leads to the "WP8" (EDG Work Package 8) group creation for common LHC Experiments activities that included also five "EU funded" persons. Their help, advisory and WP8 coordination during the Stress Test was a key element of the test successes [13].

During the test a partial set of the CMS production software/tools as described above was used. Namely the CMKIN and CMSIM steps were performed, as the combination of goals (CMS and EDG) needed only these two stages of CMS production.

Most of the involved sites (actually all of them, with different commitments) implemented the necessary Grid services, including Computing Elements (CEs), Storage Elements (SEs), Resource Brokers (RBs), Information Systems, Replica Catalogs (RCs), etc. User Interfaces (UIs) from which the CMKIN and CMSIM jobs were submitted to the Grid (via the RBs), were implemented at four CMS sites: CNAF-INFN/Bologna, Padova/INFN, Ecole Polytechnique/IN2P3 and Imperial College/London.

Different local job schedulers were also considered for the stress test, including PBS, LSF, and BQS.

As quoted above, also two different MSS systems were used for the Stress Test, HPSS at the Lyon site and Castor at the CERN site.

The resource brokers used for matching the jobs' requirements were as many as the submitting UIs, thus allowing for easy control of performances and load balance. An additional couple of RBs were also available as "backup" resources.

The used Grid middleware components included (EDG from version 1.3.4 to version 1.4.3):

- Resource Broker servers
- Replica Manager and Replica Catalog Servers
- MDS and Information Indexes Servers
- Computing Elements (CEs) and Storage Elements (SEs)
- User Interfaces (UIs)
- Virtual Organization Management Servers (VO) and Clients
- EDG Monitoring
- Software distribution via RPMs managed via LCFG.

Monitoring of the EDG CMS Stress Test was based on multiple products, allowing for redundancy and possible recover of failures. It included:

- EDG monitoring system (MDS based)
- BOSS database stored information
- Online monitoring with Nagios

Both EDG and BOSS sources were processed for post Stress Test analysis to measure performances and efficiencies of use (a special script was developed to manage and analyze the information: *boss2root*).

### 4.1. CMS-EDG Middleware integration

Figure 1 gives a picture of the implemented dependencies and integration of CMS software with the "high layer" EDG-Grid functionalities.

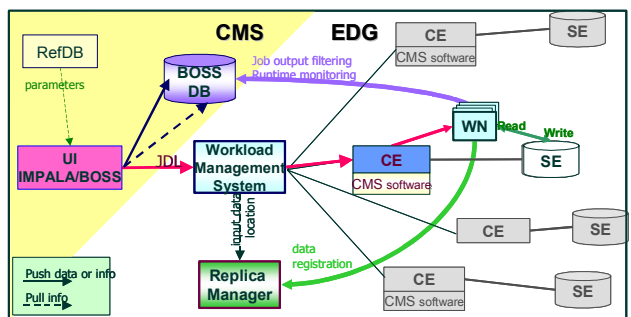

Figure 1: CMS Production tools integration with the EDG middleware components.

The CMS software (and tools) had to be adapted to interface and/or use the APIs of EDG middleware, and in particular the UIs implementations and the BOSS tool had to be modified. As shown in the Figure 1 the interface between the CMS "world" and the EDG "world" is only (mainly) confined to those two CMS components. Within the UI the IMPALA/Boss configuration had to be modified to produce "JDL" (EDG Job Description Language) aware files, which in turn were submitted to the EDG Grid Testbed via the IMPALA tool.

The BOSS tool had to be modified to cope with the distributed and Grid remote WNs, and also to deal with the additional parameters to be monitored.

The Work Load Management System (Resource Brokers) accepted the jobs and sent them to the appropriate (free and available) Computing Element (CE). The Information System (MDS) provided to the RBs the necessary information about the availability of the resources, including the location of the nearest Storage Element (SE). Matching of the job requirements, as prepared via the "JDL" on the UIs through the modified IMPALA tool, were performed on the RB, including the choice of only CEs CMS-ready (with the correct CMS software environment installed and available).

The Stress Test implementation allowed for job submission from four different UIs on the same distributed system of resources, eventually distributing automatically the computational load.

## 5. RESULTS

EDG Stress test could measure the failure and success rate of the Grid submitted CMS jobs. These measurements were possible thanks to the redundant job tracking and monitoring (EDG WMS logging and bookkeeping and CMS BOSS job tracking) performed during the test. Rates were measured for the two kinds of CMS simulation jobs submitted to the Grid: CMKIN and CMSIM.

CMKIN jobs were named "short jobs, because of their "short" CPU time requirement and light access to the Grid services. In particular the Replica Catalog was only accessed to write and register (via the Replica Manager) the final produced files.

CMSIM jobs were named "long jobs", because of their "long" CPU time requirement and heavy access to the Grid Services. CMSIM jobs need to find the input data file querying the Replica Catalog and then match the required resources via the Resource Broker. Finally the produced output had to be written to Storage and registered in the Catalog.

Tables 3 and 4 show the total number jobs submitted to EDG Testbed for both CMS simulations steps (~ 10500 jobs in about three weeks), with the breakdown of successfully finished and failed number of jobs. The tables also provide information about the measured efficiencies of jobs' successes.

A total of 6336 CMKIN jobs were launched into the Grid, and a total of 5518 were successful.

The overall efficiency during the whole Test for CMKIN jobs for EDG evaluation of the Testbed and middleware performances turned out to be 87%.

A total of 4340 CMSIM jobs were launched into the Grid, and a total of 1678 were successful.

The overall efficiency during the whole Test for CMSIM jobs for EDG evaluation of the Testbed and middleware performances turned out to be ~39%.

Further information about the performed measurements is also given in Tables 3 and 4. The second row in both tables provides information for an early 2003 (January 2003) deployed EDG middleware release, and the last row is an estimation of the same measured quantities as for the CMS experiment point of view. The CMS evaluation was performed using "only" CMS provided monitoring and tracking tools (mostly BOSS) and was intended to provide a measurement of the submitted jobs that could provide usable CMS files containing the correct simulated events (no matter what the final Grid declared status could be).

There was a clear indication of time improvement of the Grid middleware efficiency, for both kinds of computational jobs. There was also a clear indication that redundancy on job monitoring and tracking is still required to correctly identify successful jobs ("CMS tool" based evaluation).

Table 3: EDG Stress test classification of submitted jobs for the "CMKIN" like processes.

| CMKIN jobs | | | |
|---|---|---|---|
| Status | EDG Stress Test evaluation | EDG ver 1.4.3 | "CMS" Stress Test evaluation |
| Finished Correctly | 5518 | 1014 | 4742 |
| Crashed or bad status | 818 | 57 | 958 |
| Total number of jobs | 6336 | 1071 | 5700 |
| Efficiency | 0.87 | 0.95 | 0.83 |

Table 4: EDG Stress test classification of submitted jobs for the "CMSIM" like processes.

| CMSIM jobs | | | |
|---|---|---|---|
| Status | EDG Stress Test evaluation | EDG ver 1.4.3 | "CMS" Stress Test evaluation |
| Finished Correctly | 1678 | 653 | 2147 |
| Crashed or bad status | 2662 | 264 | 935 |
| Total number of jobs | 4340 | 917 | 3082 |
| Efficiency | 0.39 | 0.71 | 0.70 |

EDG Stress Test identified middleware problems and limitations, mostly due to the newly developed high-layers functionalities. Many "on the fly" solutions or corrections were implemented during the Stress test period of time, including some possible work-around.

The encountered problems mainly included:
- MDS and Information Index instability;
- Replica Catalog limitations;
- Job submission chain (Resource Broker, Job Submission Service and local scheduler interfaces) weakness related to the many underling services;
- Other sporadic Grid services unavailability or hardware failures (including Network).

The Information Index instability was due to too many accesses: the top MDS and the II slowed down dramatically once the query rate increased above a certain level and eventually hung indefinitely. Since the Resource Broker relies on the II to discover available resources, the MDS instability caused jobs to abort due to lack of matching resources.

In order to reduce the effect of the problem, a lower rate of launched jobs in the job submission process was adopted, in particular for CMSIM jobs where the RB matchmaking is more complex. Moreover a workaround solution to increase the responsiveness of the II was also adopted (since EDG version 1.4.0), replacing the II with a customized OpenLDAP server.

The Replica Catalog implementation on the EDG released software displayed some limitations when coupled with the requirements of CMS Production. Too many concurrent jobs writing into the RC overload the LDAP server, thus slowing down its performances and eventually causing it to stick. Moreover the limit of about 2000 entries (of very long character-strings identifying the file names) per catalog collection was hit.

A workaround for those limitations was adopted slowing as much as possible the job submissions and creating different Replica Catalog entries for the different UIs under the same RC.

Several problems at various levels of the job submission chain were found during the Stress Test. Identifying the reason of them was a major effort for the CMS-EDG Task Force. A partial list includes:
- jobs stuck in CondorG queue
- Logging and Bookkeeping Interlogger down
- crash of the CondorG *schedd* process due to use of too low value for some configuration parameters
- the globus-url-copy issued from the WN to the RB node, to download/upload the sandboxes files, didn't succeed ("*Failure while executing job wrapper*")
- the standard output of the script which wraps around the user job happened to be empty ("*Failure while executing job wrapper*"). Many possible reasons were identified for this kind of behavior.

Many of the identified problems were corrected during the Test and promptly implemented in the Testbed, thus reducing or even eliminating the inefficiency. Some other problems could only be corrected with a consistent revision of some services and therefore could only be applied after the end of the Stress Test (January 2003, EDG version 1.4.3 and later ones).

The systematic tracking of each job could allow for a detailed breakdown of failure reasons (and correlation of jobs failures/successes from EDG and CMS evaluations, not reported here). Tables 5 and 6 report the summarized reasons of failure for CMKIN and CMSIM jobs respectively. Classified reason of failures can be matched against identified middleware components limitations or misbehaviors, as shortly listed above in this paper.

Table 5: EDG Stress test classification of reason of failure for CMKIN jobs.

| CMKIN jobs | |
|---|---|
| **Status** | **Totals** |
| **Crashed or bad status** | **818** |
| | |
| **Reasons of Failure for Crashed jobs** | |
| No matching resource found | 509 |
| Generic Failure: MyProxyServer not found in JDL expr. | 102 |
| Running forever | 74 |
| Failure while executing job wrapper | 37 |
| Other failures | 96 |

Table 6: EDG Stress test classification of reason of failure for CMSIM jobs.

| CMSIM jobs | |
|---|---|
| **Status** | **Totals** |
| **Crashed or bad status** | **2662** |
| | |
| **Reasons of Failure for Crashed jobs** | |
| Failure while executing job wrapper | 1476 |
| No matching resource found | 722 |
| Globus Failure: Globus down/Submit to globus failed | 144 |
| Running forever | 116 |
| Globus Failure | 90 |
| Other failures | 114 |

EDG Stress test produced also ~260000 useful CMS events.

Figure 2 reports the integrated production of the final delivered CMSIM simulated events, over the period from November 30th to December 20th 2002. Periods of important meetings and holidays are clearly visible as "plateau". A special "plateau" (from 8th to 12th of December) in the integrated production is also visible and is due to the Testbed upgrade to a new and patched version of EDG middleware, during which the production had to be halted.

Peak rate of event simulation was of 2.5 seconds per event production (12th – 14th December). The average rate during the entire period was about 7 seconds per event simulation.

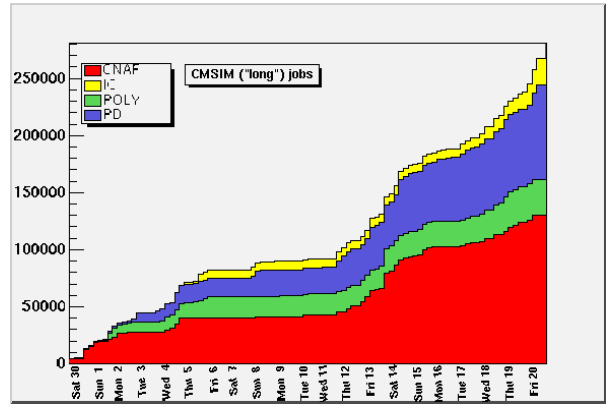

Figure 2: EDG integrated rate of CMS events production (from November 30th to December 20th, 2002).

An example of job distribution over the computing sites (CEs) is shown in Figure 3. The CE plot clearly shows the adopted strategy of job submission by the UIs and also the use of available resources.

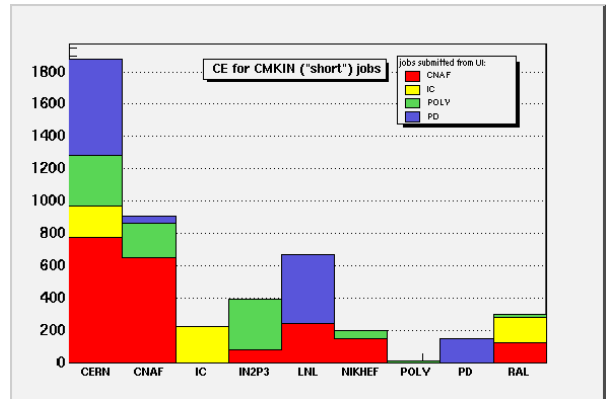

Figure 3: Number of CMKIN jobs that were executed on each CE.

## 6. CONCLUSIONS

The Stress Test of Grid environment gave many results covering different aspects of possible CMS software/tools development and Grid Projects evolution.

EDG test was focused on the measuring of Grid "higher services" performances, as well as aiming for a large amount of CMS usable simulated events.

Portability of the CMS production environment into Grid implementation was demonstrated to a high degree, giving however some good hint for possible modifications of global design architecture and/or implementation of it.

Verification of the EDG robustness in a production environment demonstrated a lack of software maturity. Though this was to be expected, given the fact that this was only the second year of the EDG R&D program, getting a robust, or at least stable, environment proved to be quite difficult during the test.

The production of CMS simulated events at the level of 1 million could not be obtained. However, more than 250000 events were produced successfully in a period of three weeks (compared to the four weeks planned). Taking into account the many interruptions of the test due to meetings, unavailability of resources and personnel, hardware failures and software changes, the result can be considered a reasonable success (even compared to "traditional" CMS productions over dedicated Farms).

Dynamic addition of new sites and resources to the testbed was demonstrated to be possible without disruption to the system as a whole, which is promising for the future as the system scales up in complexity, size and use.

Among the many lessons learn during the Stress Test, some major outcomes can be extracted and summarized as follows:
- No serious "show stopper" problem was found.
- Many bugs and limitations were found by stressing the system:
  1. Bugs (coming from many pieces of software provided by many authors) were promptly and iteratively corrected and new versions of the middleware were installed "on the fly";
  2. Limitations were correctly identified and workarounds were found in close collaboration between CMS and EDG personnel, whenever possible.
- The measured and final efficiencies (both for CMS and for EDG evaluation) were found to be:
  1. Substantially different for jobs requiring small CPU time (few seconds) and for jobs lasting longer (order of 12 hours). The range was from 95% to 40% depending also on the kind of analysis applied. The "short" jobs showed better efficiencies than the "long" ones. This can be explained by the larger complexity of "long" jobs (input/output loads, larger requests to the services, etc.);
  2. Overall, about 60% of successful jobs attained EDG completion;
  3. Overall, about 70% of successful jobs resulted in correct and available CMS files of simulated events;
  4. Major sources of inefficiencies were: Information System (MDS) instabilities, Replica Catalog performances, hardware failures and mis-configurations, fragility of the GRAM-GASS Globus mechanism for job submission and output retrieval.

A much more improved situation was experienced during the last days of the Stress Test and during the follow-up, at the beginning of 2003, when EDG release 1.4.3 was deployed. That release incorporated some Globus work-around and bug corrections for the problems quoted above. Even though a small sample of submitted jobs (~1,000) were submitted, a preliminary estimation of efficiency indicated a value of about 80% (or even better if the trivial errors are excluded) for the EDG successful jobs.

## Acknowledgments

The authors wish to thank all the CMS and EDG colleagues for invaluable help and support. Many other contributions came by other Grid projects persons, and in particular by LCG (LHC Grid Project) and EDT (European DataTag project).

This Work was supported by the CMS participating Institutions and by the European Union.

## References

[1] CMS Experiment, see for example: http://cmsdoc.cern.ch/cms/outreach/html/index.shtml
[2] LHC Project, see for example: http://lhc.web.cern.ch/lhc/general/gen_info.htm
[3] CERN (Geneva, CH), see for example: http://public.web.cern.ch/public/
[4] The CMS Production Team: T. Wildish, V. Lefebure et al.,"The Spring 2002 DAQ TDR Production", CMS Note 2002/034, Geneva, September 2002
[5] EDG Project, see for example: http://eu-datagrid.web.cern.ch/eu-datagrid/
[6] The CMS-EDG Stress-Test Task Force, "CMS Test of the European DataGrid Testbed", CMS NOTE 2003 in preparation, Geneva, May 2003.
[7] Objectivity/DB Main Web site: http://www.objectivity.com/
[8] V. Lefebure, "RefDB: A Reference Database for CMS Monte Carlo Production", CHEP03, San Diego, March 2003.
[9] C. Grandi, "BOSS: a tool for batch job monitoring and book-keeping", CHEP03, San Diego, March 2003.
[10] G. Graham, "McRunJob: A Workflow Planner for Grid Production Processing", CHEP03, San Diego, March 2003.
[11] C. Grandi, "Plans for the Integration of grid tools in the CMS computing environment", CHEP03, San Diego, March 2003.
[12] Globus Main Web site: http://www.globus.org/
[13] S. Burke et al., "HEP Applications Evaluation of the EDG Testbed and Middleware", CHEP03, San Diego, March 2003.